\documentclass{aastex631}
\usepackage{amsmath}
\usepackage{amsfonts}
\usepackage{booktabs}
\usepackage{color}
\usepackage{dblfloatfix}
\usepackage{enumerate}
\usepackage{epstopdf}
\usepackage{graphicx}
\usepackage{longtable}
\usepackage{mathptmx}
\usepackage{multirow}
\usepackage{natbib}
\usepackage{rotating}
\usepackage{xfrac}
\usepackage{hyperref}

\begin{document}
\title{Two repeated quasi-periodic oscillations in the FSRQ S5 1044+71 observed by TESS}

\author{Jun-Jie Wang}
\affiliation{Key Laboratory of Colleges and Universities in Yunnan Province for High-energy Astrophysics, Department of Physics, Yunnan Normal University, Kunming 650500, China}
\affiliation{Yunnan Province China-Malaysia HF-VHF Advanced Radio, Astronomy Technology International Joint Laboratory, Kunming 650011, China}
\affiliation{Guangxi Key Laboratory for the Relativistic Astrophysics, Nanning 530004, China}
\author{Ting-Feng Yi}
\affiliation{Key Laboratory of Colleges and Universities in Yunnan Province for High-energy Astrophysics, Department of Physics, Yunnan Normal University, Kunming 650500, China}
\affiliation{Yunnan Province China-Malaysia HF-VHF Advanced Radio, Astronomy Technology International Joint Laboratory, Kunming 650011, China}
\affiliation{Guangxi Key Laboratory for the Relativistic Astrophysics, Nanning 530004, China}
\correspondingauthor{Tingfeng Yi; Yangwei Zhang}
\email{yitingfeng@ynnu.edu.cn; zhangyangwei@ynu.edu.cn}
\author{Yangwei Zhang}
\affiliation{South-Western Institute for Astronomy Research, Yunnan University, Kunming 650091, China}
\author{He Lu}
\affiliation{Key Laboratory of Colleges and Universities in Yunnan Province for High-energy Astrophysics, Department of Physics, Yunnan Normal University, Kunming 650500, China}
\author{Yuncai Shen}
\affiliation{Key Laboratory of Colleges and Universities in Yunnan Province for High-energy Astrophysics, Department of Physics, Yunnan Normal University, Kunming 650500, China}
\author{Lisheng Mao}
\affiliation{Key Laboratory of Colleges and Universities in Yunnan Province for High-energy Astrophysics, Department of Physics, Yunnan Normal University, Kunming 650500, China}
\author{Liang Dong}
\affiliation{Yunnan Province China-Malaysia HF-VHF Advanced Radio, Astronomy Technology International Joint Laboratory, Yunnan Observatories, Chinese Academy of Sciences, Kunming 650011, China}
\begin{abstract}
In this work, we report for the first time two repeated quasi-periodic oscillations (QPOs) in the light curve of the Flat Spectrum Radio Quasar (FSRQ) S5 1044+71. This source was observed by the Transiting Exoplanet Survey Satellite (TESS) in multiple sectors. We used the generalized Lomb-Scargle periodogram method and weighted wavelet Z-transform method to search for significant periodic signals. The main results are as follows: We found QPOs of $\sim$ 7.0 days (persisted for 4 cycles, with a significance of $\sim3.5\sigma$) and $\sim$ 7.3 days (persisted for 5 cycles, with a significance of $\sim3.8\sigma$) in the light curves of Sector 47 and EP1, respectively. Considering range of error, we consider them to be the same. We discussed two likely models of these rapid quasi-periodic variations: One comes from the jet and the other from the accretion disk. For the first one, we consider kink instability of the jet as a plausible explanation. Second, the QPO is probable to come from the main hot spots in the accretion disk, which is located approximately within the innermost stable circular orbit allowed by general relativity. Based on this model, we estimate the mass of the black hole in S5 1044+71 to be $3.49 \times 10^9 M_{\odot}$.
\end{abstract}


 \section{Introduction}           
\label{sect:intro}

Active Galactic Nuclei (AGN) are a special class of objects with intense activity or violent physical processes which mainly occurring in or triggered by the cores of galaxies. Blazars are a subclass of AGN characterized by highly relativistic speeds and directed towards Earth (with jets at angles equal to or less than 10°) \citep{1995PASP..107..803U}, exhibiting extreme luminosity and motion characteristics, all originating from the galaxy's nucleus \citep{2019MNRAS.484.5785G}. Blazars can be divided into two types: BL Lacertae objects (BL Lac) and Flat Spectrum Radio Quasars (FSRQ). BL Lac objects show featureless continuum spectra or extremely faint emission lines \citep{1991ApJS...76..813S}, while FSRQs exhibit prominent emission lines in the optical and ultraviolet bands \citep{1978PhyS...17..265B}. Blazars are among the most variable extra-galactic objects, showing significant rapid variability in multiple bands, with high non-thermal radiation \citep{2019MNRAS.484.5785G}.  

Quasi-Periodic Oscillations (QPOs) sometimes occur in blazars. QPOs have been reported across the entire electromagnetic spectrum in blazars, from radio to gamma-rays, with diverse timescales ranging from minutes to days \citep{2014JApA...35..307G}, weeks \citep{2010fym..confP..39G}, and even years \citep{2009ATel.2353....1G, 2016AJ....151...54S, 2022ApJ...929..130W}. QPOs may originate within a few gravitational radii of black holes, making them useful for studying the gravitational influence of central objects on their surrounding environments. The physical origin of these QPOs can be explained by the various models, including disk seismology, warped accretion disks, or disk-jet coupling  \citep{1999ApJ...524L..63S, 2021BlgAJ..34..103T, 2001ApJ...559L..25W, 2005PASJ...57..699K}. The latest research by \cite{2022ApJ...933...55C}, based on three-dimensional general relativistic magnetohydrodynamic simulations of accretion flows, suggests that quasi-periodic jets may be generated by magnetic reconnection within the flow forming flux ropes. The QPO value associated with the accretion disk is proportional to the mass of the central object. Despite these efforts, the physical mechanism behind the origin of QPOs remains elusive.

Due to the uneven and irregular sampling of ground-based measurements, optical QPOs are difficultly detected. When searching for periodicity in this kind of data, quasi-periodic signals from AGNs may be disrupted by the characteristics of random red noise, potentially being masked  \citep{2016MNRAS.461.3145V}. The Transiting Exoplanet Survey Satellite (TESS) has largely overcome these issues by improving telescope precision, enabling high-cadence regular sampling. TESS is a space-based optical instrument, and can generate high-precision, high-cadence light curves. Unlike ground-based telescopes which are affected by seasonal gaps, irregular sampling, and atmospheric-induced photometric accuracy degradation. This study focuses on the AGN sample set detected in TESS observations, with a total of 176 identified AGN from the first sector to the 76th sector. By visual inspect of the light curves for each source in this sample set, we discovered a new optical QPO possibly existing in S5 1044+71. 

S5 1044+71 belongs to FSRQ in the classification of blazars, the redshift is 1.14. and it was previously discovered to have a long-term QPO of about 3 years in the gamma-ray band \citep{2022ApJ...929..130W}. Interestingly, in this paper, the discovered QPO that we find in the TESS optical band light curve is a short-term QPO of about 7 days. We summarize basic information about the TESS satellite and its instrument particulars in Section 2. In Section 3, we present the TESS observations and data reduction of the FSRQ S5 1044+71. In Section 4, we brieﬂy describe data analysis and give the results of these analyses, indicating the presence of a repeated QPO. A discussion and conclusions are in Section 5.


\section{Instrument Particulars}

TESS is a space telescope launched by the National Aeronautics and Space Administration (NASA) in April 2018, with the primary goal of searching for exoplanets orbiting the brightest dwarf stars. TESS is equipped with four wide-field optical Charge-Coupled Device (CCD) cameras, each with a Field-of-View (FOV) of 24° × 24°. These four cameras are linked together to form a 24° × 96° field of view. TESS divides the celestial sphere into 26 strip-like regions called “sectors”, and there are 13 sectors in the northern and southern celestial hemispheres, respectively. The observation range of one sector is defined by the field of view of the four CCD cameras. It takes approximately 27 days to complete the observation of one sector, and extrapolating this, it would take about two years to complete a full-sky survey \citep{2018cosp...42E2851R}. Due to the rectangular field of view of TESS, there are specific regions in the sky that fall into multiple sectors, especially near the poles, leading to coverage gaps near the celestial equator \citep{2023ApJ...943...53K}. TESS has its own data repository containing observations of target celestial bodies in the optical band. Data collected by TESS can be accessed from the Mikulski Archive for Space Telescopes (MAST) database \citep{2019AJ....157...98G}. Additionally, the Python file ``Quaver.py'' mentioned in Krista Lynne Smith's article \citep{2023ApJ...958..188S} can be used to extract data, no need for us to manually extract from the website, simply inputting the TESS name of the source will automatically extract.
\begin{figure*}[h]
    \centering
    \includegraphics[width=1\textwidth]{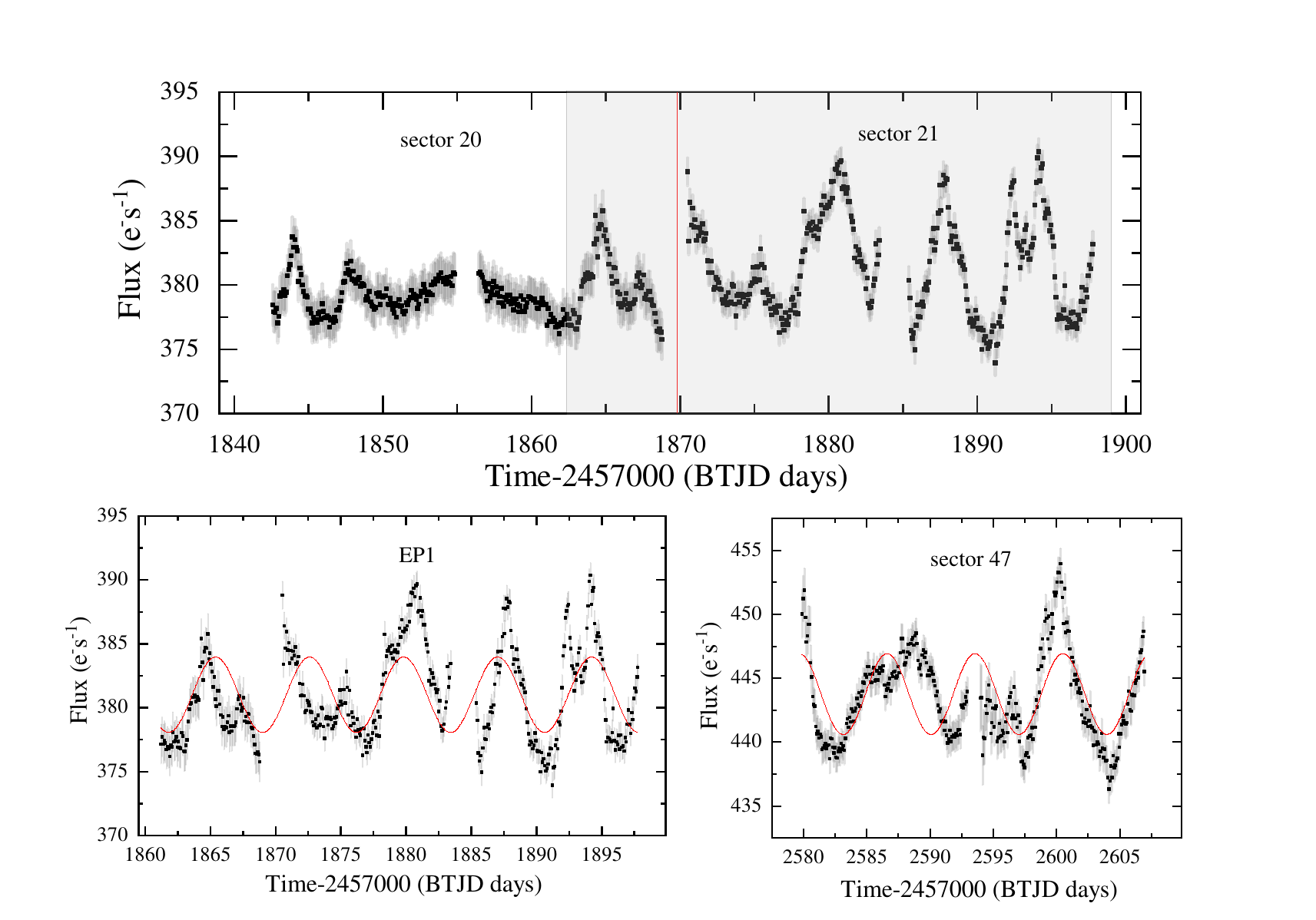}
    \caption{The light curves (two-hour binned) of S5 1044+71 for sectors 20, 21, EP1, and 47 are shown in the figure. The black points represent the actual data points (flux), while the gray indicates the associated errors.The upper half of the figure shows the combined light curves of sectors 20 and 21, where the red lines on the left and right represent the light curves for sectors 20 and 21, respectively. The gray rectangular area represents the EP1 data extracted from the merged sector light curve. The light curve for EP1 is shown in the lower left panel, revealing five complete peaks (indicated by the red sinusoidal line). The light curve for EP1 is shown in the lower right panel, revealing four complete troughs (indicated by the red sinusoidal line).}
\end{figure*}

\section{Observations and Data Reduction}
To ensure the determination of our sample for analysis, we downloaded all observed targets in 76 sectors from the link \footnote{\url{https://archive.stsci.edu/tess/bulk_downloads/bulk_downloads_ffi-tp-lc-dv.html}} of the TESS official website. Most of the targets are stellar, with only a few being non-stellar objects. Through an our Python script, we filtered out 176 source from all the targets. We utilized the Quaver code for light curve extraction and systematic calibration. The convenience of this software lies in the ability to custom-select extraction regions from the cutouts of full-frame images (FFIs) without having to download the entire image, effectively avoiding contamination from nearby sources \citep{2019ascl.soft05007B}. Furthermore, Quaver code allows for fitting of the calibrated target light curves using three methods: Full hybrid overfit (FHO), Simple hybrid overfit (SHO) and Simple PCA overfit (SPO). Importantly, the Quaver code \citep{2023ApJ...958..188S} is open-source and comes with detailed usage instructions\footnote{\url{https://github.com/kristalynnesmith/quaver}}. And we finally decided to use the SHO method to fit  all the data of FSRQ S5 1044+71 observed by TESS.

To performed a detailed time series analysis of sources, it is necessary to perform special processing on the light curves from bright host galaxies and surrounding resolved sources. This process involves constructing a reasonable background variability model and employing principal component analysis at different stages to address more subtle multiplicative systematics. Finally, linear regression is used to remove these effects to avoid overfitting the original light curves of the AGN. First, we download 25 pixels × 25 pixels TESS full-frame “postage stamp" images centered on the target. Quaver code has provided us with a customized aperture that can mitigate the influence of surrounding sources. Subsequently, using the matrix regression method available in the publicly accessible TESSCUT and LIGHTKURVE software packages \citep{2012ASPC..461..627A} , we fit the faint background pixels as the dominant systematic errors in the light curve. Finally, the original light curve undergoes background systematics correction to obtain the processed light curve. The light curves of the FSRQ S5 1044+71 extracted from the observational data of TESS sector 14, 20, 21, 40, 41, 47, 48, 60, 74, and 75 are shown in Figure 1 and Figure 2.Differences in the average flow of the source in different sectors may be caused by variations in the angle of satellite observations, in addition to variations in the state of the source itself. Due to the continuity and similar waveform between the latter part of sector 20 (BTJD 2458860-2458869 day) and sector 21, we combined them into a single time region, denoted as Epoch 1 (BTJD 2458860-2458898 day), abbreviated as EP1. The difference in average flux between the two sectors is small but negligible relative to the overall intensity. This QPO we found is a transient QPO, not a permanent QPO, so we analyze it using partial light curves rather than overall light curves. 

\begin{figure*}[h]
    \centering
    \includegraphics[width=1\textwidth]{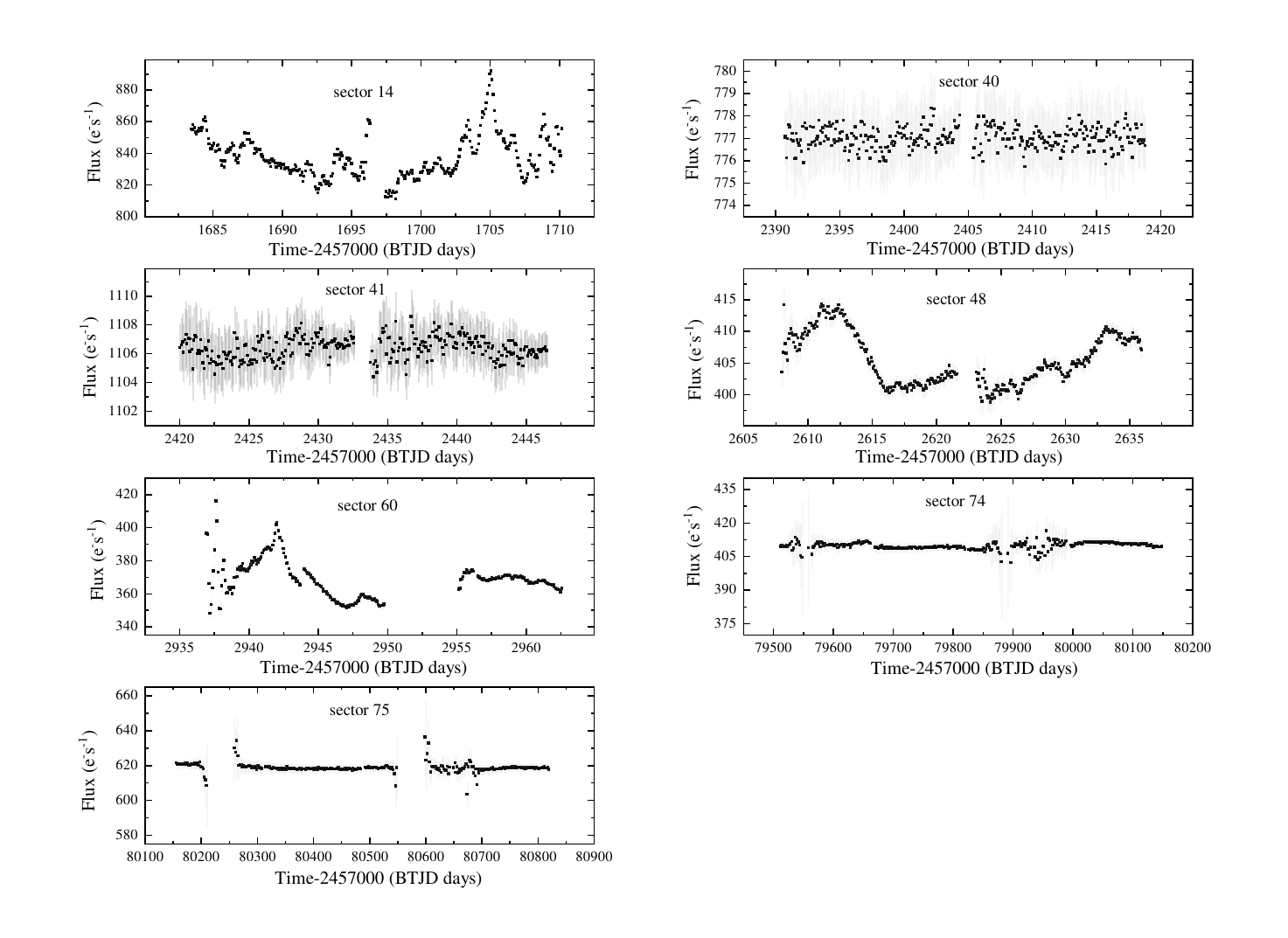}
    \caption{The light curves (two-hour binned) of S5 1044+71 observed in 7 sectors, namely sectors 14, 40, 41, 48, 60, 74 and 75. The black points represent the actual data points (flux), while the gray indicates the associated errors.}
\end{figure*}
\section{Data Analysis and Results}
\subsection{Data Segmentation and Binning}
The gap of approximately one day in the processed light curve may be due to TESS transmitting data to Earth or Earth sending commands to TESS. Due to the large number of data points, for ease of subsequent data processing, we binned the data by averaging the data points within two hour and treating them as a single point. The difference in average values of light curves for different sectors may be due to the varying observational angles of S5 1044+71 in each sector.
\subsection{Lomb–Scargle Periodogram (LSP)}
The periodogram of a time series can be used to identify periodicity through the discrete Fourier transform. The Lomb-Scargle Periodogram (LSP) is a useful Fourier transformation method commonly employed for analyzing the periodicity of irregularly and unevenly sampled data  \citep{1976Ap&SS..39..447L, 1982ApJ...263..835S}. LSP uses the $\chi^2$ statistic to fit sine and cosine functions to the entire time series data. LSP fits a sine model to the data for each frequency, and if the power at a particular frequency is high, then that frequency is likely to exist. Assuming a variance of $\sigma^2$, the expression for LSP  \citep{1986ApJ...302..757H} is given as
\begin{equation}
P_{L S}(\omega)=\frac{1}{2}\left\{\frac{\sum_{i=1}^N\left(x\left(t_i\right)-\bar{x}\right) \cos \omega\left(t_i-\tau\right)}{\sum_{i=1}^N \cos ^2 \omega\left(t_i-\tau\right)}+\frac{\sum_{i=1}^N\left(x\left(t_i\right)-\bar{x}\right) \sin \omega\left(t_i-\tau\right)}{\sum_{i=1}^N \sin ^2 \omega\left(t_i-\tau\right)}\right\},
\end{equation}
where, $t_i$ is the measurement time; $x_i$ is the corresponding flow value; $\bar{x}$ is the average value of $x_i$; $\omega$ is the frequency; $\tau$ is the phase correction for the corresponding time $t$
\begin{equation}
\tau=\frac{1}{4 \pi f} \tan ^{-1}\left(\frac{\sum_n \sin \left(4 \pi f t_n\right)}{\sum_n \cos \left(4 \pi f t_n\right)}\right).
\end{equation}

\subsection{Weighted wavelet Z-transform analysis}
Astronomical signals often exhibit short-term characteristic oscillations rather than constant amplitude and phase \citep{2023ApJ...943...53K}. In such cases, Fourier analysis may not be the optimal solution. The wavelet transformation method focuses on the short-term and limited time span of data, enabling the detection of any possible periodicity in the signal and an evaluation of its persistence throughout the entire observation process. Wavelet transformation simultaneously decomposes the signal into frequency and time domains. This method uses wavelet functions to fit the observed values instead of sinusoidal components, making it highly suitable for detecting transient periodic oscillations. Wavelet Analysis evolved from Fourier Transform and it can examine the time evolution of parameters (period, amplitude and phase) describing periodic and quasi-periodic signals. According to the complex Morlet wavelet function:
\begin{equation}
\psi(t)=e^{-\frac{t^2}{2}\left(e^{i \omega_0 t}-e^{-\omega_0 / 2}\right)},
\end{equation}
where $\omega_0$ represents the attenuation factor.
\begin{equation}
\psi\left(\frac{t-b}{a}\right)=e^{-\frac{(t-b)^2}{2 a^2}} e^{i \omega_0\left(\frac{t-b}{a}\right)},
\end{equation}
where {{$a$} and {$b$} represent the scaling and translation parameters, respectively. Then, Foster defines weighted wavelet transform ({WWT}) \citep{1996AJ....112.1709F}, 
\begin{equation}
W W T=\frac{\left(N_{e f f}-1\right) V_y}{2 V_x},
\end{equation}
\begin{equation}
N_{eff}=\frac{\left(\sum \omega_\alpha\right)^2}{\sum \omega_\alpha^2},
\end{equation}
\begin{equation}
V_x=\frac{\sum_\alpha \omega_\alpha x^2 t_\alpha}{\sum_\beta \omega_\beta}-\left[\frac{\sum_\alpha \omega_\alpha x t_\alpha}{\sum_\beta \omega_\beta}\right]^2, V_y=\frac{\sum_\alpha \omega_\alpha y^2 t_\alpha}{\sum_\beta \omega_\beta}-\left[\frac{\sum_\alpha \omega_\alpha y t_\alpha}{\sum_\beta \omega_\beta}\right]^2, 
\end{equation}}
where $N_{eff}$ is the effective number of data points; $V_x$ is the weighted deviation of the data; $V_y$ is the weighted deviation of the model function. This principle states that at lower frequencies, the effective data $N_{eff}$ is greater than at higher frequencies, so the values of the weighted wavelet transform are biased towards higher frequencies. Therefore, when incorporating the Z-statistic, we refer to it as the Weighted Wavelet Z-Transform(WWZ):
\begin{equation}
Z=\frac{\left(N_{\text {eff }}-3\right) V_y}{2\left(V_x-V_y\right)}.
\end{equation}

\subsection{Light Curve Simulation and Peak Significance}
The peaks at lower frequencies in the power spectrum  can be mistaken for quasi-periodic signals. Therefore, the evaluation of the significance level of these peaks is a crucial step. We simulate the artificial light curve using the “DELightcurveSimulation" code given by \cite{2016ascl.soft02012C} . The code is a Python implementation of the Mathematica code provided by \cite{2013MNRAS.433..907E}. The simulated light curves have the same variability and statistical properties as the observed light curves. Based on the above method,  we simulated 10000 artificial light curves with the same PSD and PDF distributions as the original light curves.We fit the PSD and identify specific features by a simple power-law model with the following formula \citep{1995A&A...300..707T}:

\begin{equation}
P(f) \propto f^{-\beta},
\end{equation}
The red noise level $\beta$ we take the value of 1 \citep{2020PASP..132d4101Y}. For evaluating the confidence level, we calculate the WWZ power and LSP power of each artificial light curve, and calculate confidence level power, and finally draw confidence level contour lines in the WWZ power map and LSP power map, respectively.

 \subsection{Results}
\begin{figure*}[h]
    \centering
    \includegraphics[width=1\textwidth]{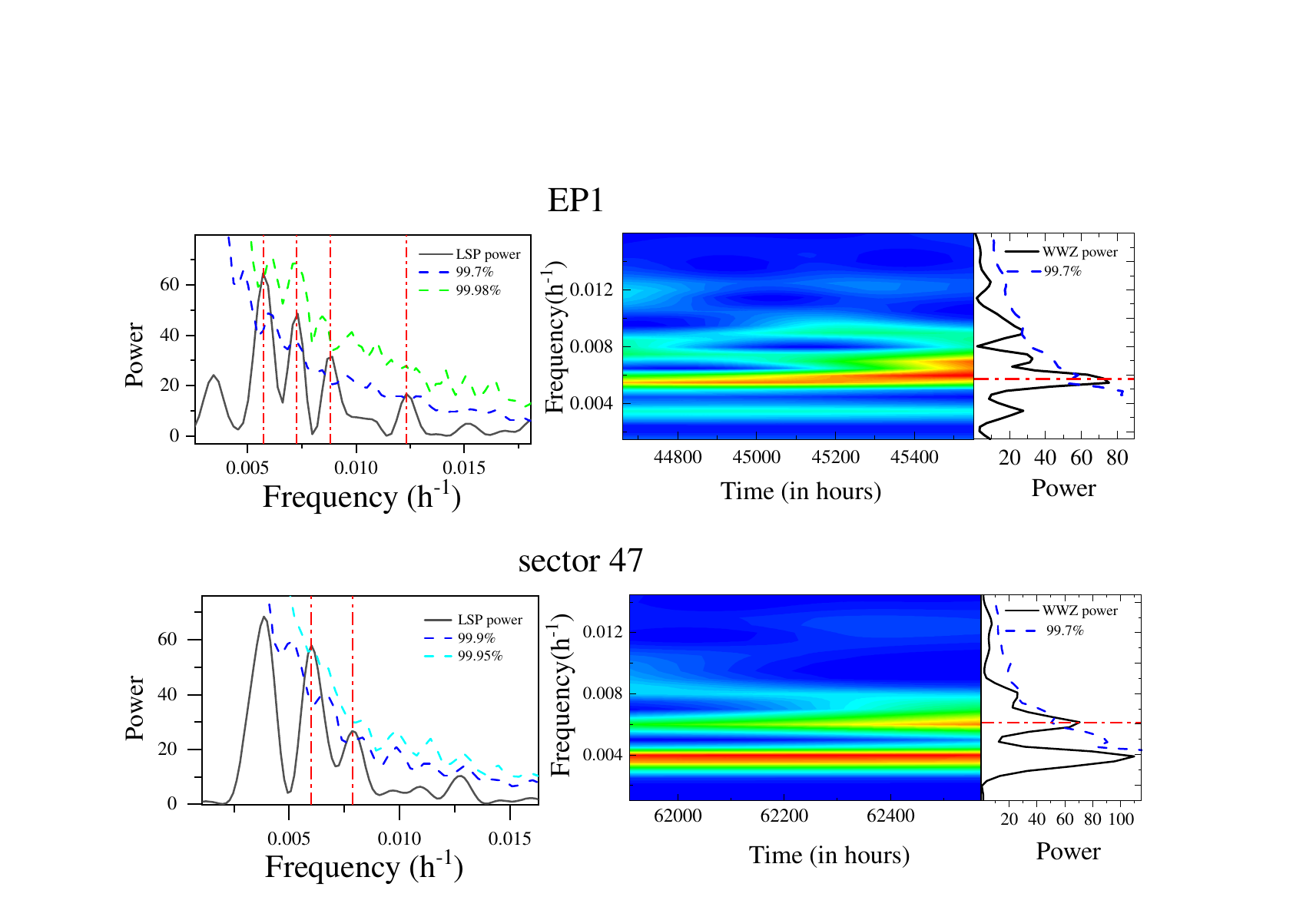}
    \caption{The results of LSP and WWZ methods for Sectors EP1 and 47.   The left panels show the LSP power (black solid line) , the red noise (red solid line)and confidence level (blue dash line and green dash line); The right panels show WWZ results. In EP1, the LSP method shows four peaks exceeding $3\sigma$. Obviously, the peak at frequency of 0.005731 \text{h\textsuperscript{-1}} ($\sim$ 7.3 days) even reach $3.8\sigma$. However, in the WWZ method, only the peak at 0.005731\text{h\textsuperscript{-1}}  exceeds $3\sigma$ in EP1. Coincidentally, sector 47 also shows two peaks exceeding $3\sigma$ in the LSP method. Similarly, in the WWZ method, only the peak at 0.005992 \text{h\textsuperscript{-1}} ($\sim$ 7.0 days) exceeds $3\sigma$ in sector 47. Obviously, two similar QPOs appear in EP1 and sector 47, respectively.}
\end{figure*}
\begin{figure*}[h]
    \centering
    \includegraphics[width=0.8\textwidth]{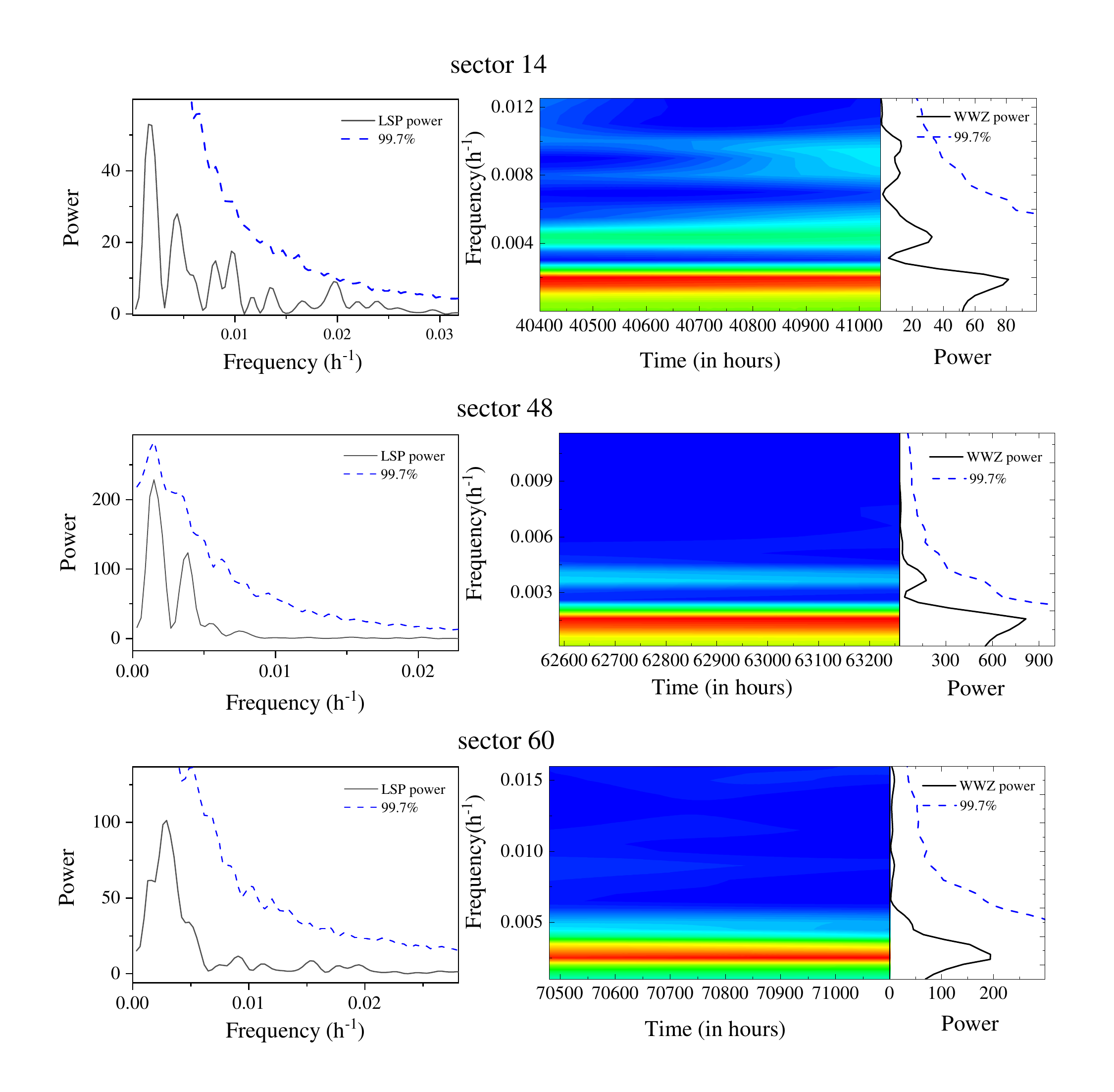}
    \caption{The results of LSP and WWZ methods for Sectors 14 , 48 and 60. The left panels show the LSP power (black solid line) and confidence level (blue dash line); The right panels show WWZ results.}
\end{figure*}
We used the LSP and WWZ methods to examine the light curves with significant variability for the presence of QPO. And we detected four peaks in the EP1 with frequencies of 0.005731 \text{h\textsuperscript{-1}} , 0.007270 \text{h\textsuperscript{-1}} , 0.008799 \text{h\textsuperscript{-1}}  and 0.012328 \text{h\textsuperscript{-1}}, using the LSP method, as shown in the upper left panel of Figure 3, with significances all $>3\sigma$, and the peak of frequency  0.005731 \text{h\textsuperscript{-1}}  even reaching $3.8\sigma$. Similarly, using the LSP method, in sector 47, we detected two strong QPO signals with frequencies of 0.005992 \text{h\textsuperscript{-1}}  and 0.007994 \text{h\textsuperscript{-1}} , as shown in the lower left panel of Figure 3, with significances $>3.0\sigma$. The peak of frequency  0.005992 \text{h\textsuperscript{-1}} even reaching $3.5\sigma$. WWZ analysis of the light curves of EP1 and sector 47 shows that these signals are persistent throughout the entire observation period. The color density plots of the WWZ power indicate that only the frequencies of 0.005731 \text{h\textsuperscript{-1}} ($\sim$ 7.3 days) and 0.005992 \text{h\textsuperscript{-1}} ($\sim$ 7.0 days) have significances$>$3$\sigma$, as shown in the right panel of Figure 3. We take the half width at half-maximum (HWHM) of the peak as the period error (i.e. 7.3$\pm$0.9 days and 7.0$\pm$1.1 days), so we consider that they are the same within the error range.
Given that the more cycles there are, the higher its credibility \citep{1992A&A...264...32K,2008AJ....135.2212X}. The 7.3 days QPO displays five cycles, and the 7.0 days QPO displays four cycles, so we believe the possibility of originating from red noise is very small \citep{2016MNRAS.461.3145V}. 
The results of the analysis for sectors 14, 48 and 60 are shown in Figure 4. The top portion of the figure shows the LSP and WWZ analysis results for sector 14. In the LSP plot, the highest peak with a frequency of 0.0197 \text{h\textsuperscript{-1}}($\sim$ 2.1 days) is not reach $3\sigma$, and no peaks close to $3\sigma$ are found in the WWZ plot. The middle portion of the graph shows the LSP and WWZ analysis results for sector 48, where the highest peak at 0.00155 \text{h\textsuperscript{-1}}($\sim$ 26 days) is not reach $3\sigma$ in both the LSP and WWZ plots. The lower part of the graph shows the results of the LSP and WWZ analyses for sector 60, where there are no peaks close to $3\sigma$ in either the LSP or WWZ graph. As shown in Figure 2, the light curves in sectors 40, 41, 74, and 75 we did not observe obvious variability in them, they have a low QPO confidence level. Therefore their LSP and WWZ are not shown in this paper. Since the $3\sigma$ is our minimum criterion for interest in these peaks, we did not further analyze the aforementioned sectors.

The light curves analyzed in this work have two segments separated by wide gaps of about 2 days. However, the duration of the light curves for EP1 and Sector 47 was 35 and 27 days, respectively. So the $\sim$ 2 days gap is negligible in comparison to the duration of light curves for EP1 and sector 47  \citep{2024MNRAS.527.9132T}.

\section{Discussion and Conclusion}
In this work, we used “Quaver" code to perform preliminary analysis on the data from 10 sectors of the blazar S5 1044+71 measured by the TESS satellite, and discovered a potential periodicity in EP1 and sector 47. We calculated the possible QPO of $\sim$ 7 days using two different methods of LSP and WWZ. The optical radiation from the blazar originates from the core region \citep{2001ApJS..134..181J}, exhibiting strong variability  \citep{2016ApJ...824L..20A, 2017A&A...603A..25A} ranging from days to years  \citep{2010MNRAS.402.2087V, 2014APh....54....1A}. However, the peaks in the power spectra of the light curve of blazar are predominantly red noise, with only a small fraction showing statistically significant QPOs  \citep{1998ApJ...504L..71H, 2015Natur.518...74G, 2001ApJS..134..181J, 2024MNRAS.527.9132T}. The mechanism of the QPO is still unclear, there are some models attempting to explain this phenomenon.

Several models based on supermassive black hole binary (SMBHB) systems \citep[e.g.][]{2008DDA....39.1505V,2010MNRAS.402.2087V,2015ApJ...813L..41A} , such as continuous jet oscillation of the accretion disk and lensing-Thirring precession \citep[e.g.][]{1998ApJ...492L..59S,2000A&A...360...57R,2018MNRAS.474L..81L} , have been proposed to explain the years-long periodic QPOs observed in blazars. For the short-term QPO, there are two main explanations: turbulent model and Kink instability model. 

First, especially in the optical band of FSRQ, the QPO could potentially arise from transient oscillations in various local regions of the turbulent in accretion disk, i.e. changes in the position of hotpots and pulsation modes captured within the disk  \citep{1997ApJ...476..589P,  2005AN....326..782A, 2008ApJ...679..182E}. The location of the hot spot is approximately within the innermost stable circular orbit allowed by general relativity.Based on this model, we use the expression proposed in article of \cite{2009ApJ...690..216G}
\begin{equation}
\frac{M_{\mathrm{BH}}}{\mathrm{M}_{\odot}}=\frac{3.23\times 10^4P}{\left(r^{3/2}+a\right)(1+z)},
\end{equation}
to estimate the mass of the black hole, where $M_{\odot}$ is the mass of the sun; $P=7.3$ is the period in seconds; $z = 1.14$ is redshift; $a$ is the angular momentum parameter; $r$ is the radial distance, which is set to the innermost stable circular orbit (ISCO). In the extreme Kerr black hole model, $r=1.2$, $a=0.9982$  \citep{2009ApJ...690..216G,2008ApJ...679..182E}. Thus, we estimate the black hole mass of FSRQ S5 1044+71 to be $3.49 \times 10^9 M_{\odot}$ (extreme Kerr black hole). We consider this short-term QPO ($\sim$ 7days) in octical light curves of S5 1044+71 is possibly due to the disk of primary black hole of the binary supermassive black hole system, as it is widely believed that there is a binary supermassive black hole present in the system \citep{2023A&A...672A..86R,2022ApJ...929..130W}. In some other possible models, the QPO maybe related to the jet.  For the long term QPO, the periodic variations in Doppler boosting due to jet precession may generate QPO signals  \citep{1992A&A...259..109G}.
The cause of this jet precession is attributed to the Lense-Thirring effect \citep{1998ApJ...492L..59S, 2000A&A...360...57R, 2004ApJ...615L...5R, 2018MNRAS.474L..81L}. For short-term QPO, we consider that the kink instability is also a possible model \citep{2020MNRAS.494.1817D}. In this model, the periodicity arises from plasma instabilities driven by turbulence, referred to as the kink instability. Relativistic magneto-hydrodynamic simulations indicate that the kink instability occurs in jets with strong toroidal magnetic fields, leading to distortion of field lines and increased particle acceleration \citep{2009ApJ...700..684M, 2017MNRAS.469.4957B, 2024MNRAS.527.9132T}. The twisting of the toroidal magnetic field and the resulting particle acceleration lead to the formation of a quasi-periodic knot along the jet direction. Quasi-periodic kinks appear in jets with growing distorted magnetic fields, accompanied by a compressed, moving plasma region, forming an observable signal with QPO characteristics. 
As the plasma region in the magnetized jet is compressed during its evolution over time, such periodic kinks can exhibit quasi-periodic features that can be detected on time scales of days to weeks.\citep[e.g.][]{2022Natur.609..265J, 2017MNRAS.469.4957B}. 
According to the equation proposed by \cite{2020MNRAS.494.1817D}, 

\begin{equation}
T_{\mathrm{obs}} = \frac{R_{\mathrm{KI}}}{\langle v_{\mathrm{tr}} \rangle \delta}
\end{equation}
We can calculate the kink growth time ($\tau_{\mathrm{KI}}$) by using the ratio of the lateral displacement of the jet from its center ($R_{\mathrm{KI}}$) to the mean velocity of motion ($v_{\mathrm{tr}}$). Here, $\delta \sim 10$ is the Doppler factor, and the $R_{\mathrm{KI}}$ of a typical emission region of a flare is about $10^{16}-10^{17}$cm, and $v_{\mathrm{tr}}$ is about 0.16$c$. Thus, the estimate of 2.4-24 days for $T_{\mathrm{obs}}$ is just within the range of QPO claimed by the observations analyzed in this study. Thus, the QPO may also be due to the growth of kink instabilities in relativistic jets. 
Through visual inspection, in the lower-left panel of Fig.1, there seem to be two alternating signals in the light curve, the first one is the main signal and the second one is a weak signal. The two signals seem to have a consistent periodicity. The bimodal structure in the light curve of OJ 287 is similar to our light curve, and to study the bimodal structure in OJ 287 \cite{1998MNRAS.293L..13V} proposed the double-jet model based on SMBHB systems. The motion of the jet ejected from one of the two black holes causes the main periodicity of the light curve, while the jet ejected from the other black hole also causes the weak signal \citep[e.g.][]{2018ApJ...854...11T, 2019ApJ...875L..22C}.However, the orbital period of the SMBHB system is generally around years, which does not match the period of days we have measured. For this type of short-term dual signal, we can also consider another possible explanation, which is the existence of two hot spots on the innermost stable circular orbit of the accretion disk. The physical origin of the QPOs of blazars is still hugely debated. A large number of high-cadence and long-term high-quality observations are still needed to solve this problem.

\section*{Acknowledgments}
This research was supported by the National Natural Science Foundation of China (Grant Nos.: 12203041, 12063007, 11863007) and Yunnan Province China-Malaysia HF-VHF Advanced Radio Astronomy Technology International Joint Laboratory (Nos. 202303AP140003). This study has made use of the TESS data, obtained from the Mikulski Archive for Space Telescopes (MAST), provided by the NASA Explorer Program.

\end{document}